\documentclass[aps,prd,showpacs,preprint,groupedaddress]{revtex4}

\usepackage{graphicx}

\begin{document}

\preprint{IFIC/05-22}

\title{ULTRA HIGH ENERGY TAU NEUTRINOS AND FLUORESCENCE DETECTORS: A
  PHENOMENOLOGICAL APPROACH}

\author{M. M. Guzzo}
\email[]{guzzo@ifi.unicamp.br}

\author{C. A. Moura Jr.}
\email[]{moura@ifi.unicamp.br}
\altaffiliation[Also at ]{Instituto de F\'{\i}sica Corpuscular -
  C.S.I.C./Universitat de Val\`encia, Spain}
\affiliation{Instituto de F\'{\i}sica ``Gleb Wataghin'' - UNICAMP\\
             13083-970 Campinas SP, Brazil}

\date{\today}

\begin{abstract}
We investigate the possibility of detecting ultra-high energy cosmic
tau-neutrinos by means of a process involving a double extensive air
shower, the so-called Double-Bang Phenomenon. In this process a
primary tau-neutrino interacts with an atmospheric quark creating a
hadronic extensive air shower that contains a tau which subsequently
decays creating a second extensive air shower. The number of these
events strongly depends on the cross section and on the flux of
ultra-high energy tau-neutrinos arriving at the Earth's atmosphere.
We estimate the potential of optical detectors to observe Double-Bang
events induced by tau-neutrinos with energies of about 1~EeV whose
detection may confirm the maximal mixing observed in the atmospheric
neutrinos also for ultra-high energy neutrinos, and give information
on the neutrino flux and cross-section. For neutrino-nucleon Standard
Model extrapolated cross-section and thick source model of flux (MPR),
we estimate an event rate of 0.48 yr$^{-1}$ for an observatory with
two fluorescence detectors with 90\% efficiency in the neutrino energy
range $0.5<E_{\nu}<5$~EeV.
\end{abstract}


\maketitle

\section{\label{intro}Introduction}
It is believed that ultra-high energy (UHE) cosmic neutrinos may play
an important role to explain the origin of cosmic rays with energies
beyond the GZK limit of few times $10^{19}~$eV~\cite{G,ZK}, once that
neutrinos hardly interact with cosmic microwave background or
intergalactic magnetic fields, keeping therefore its original energy
and direction of propagation. Even if they have masses or magnetic
moments, or travel distances of the order of the visible universe,
those characteristics do not change very much.  Possible sources of
these UHE neutrinos, like Active Galactic Nuclei and Gamma Ray Bursts,
are typically located at thousands of Mpc~\cite{HSaltzberg,Halzen}.

Considering that neutrinos come from pions produced via the process
$\gamma + p \rightarrow N + \pi$ \cite{Halzen}, that there is an
additional $\nu_e$ flux due to escaping neutrons and that about 10\%
of the neutrino flux is due to proton-proton ($pp$) interactions, the
proportionality of different neutrino flavors result:
$\nu_e:\nu_{\mu}:\nu_{\tau}=0.6:1.0:<0.01$~\cite{learned}.
Nevertheless, observations of solar~\cite{solar} and
atmospheric~\cite{atmospheric} neutrinos present compelling evidence
of neutrino flavor oscillations. Such oscillations have been
independently confirmed by terrestrial
experiments. KamLAND~\cite{kamland} observed $\bar{\nu}_e$
disappearance confirming (assuming CPT invariance) what has been seen
in solar neutrino detections and K2K~\cite{ktokplb,ktokprl} observed
$\nu_{\mu}/\bar{\nu}_{\mu}$ conversion compatible with what has been
detected in atmospheric neutrino observations.

In order to understand these experimental results by means of neutrino
oscillations, two scales of mass squared differences and large mixing
angles have to be invoked. For solar and KamLAND observations, $\Delta
m^2_{\odot}\sim7\times10^{-5}~$eV$^2$ and
$\sin^22\theta_{\odot}\sim0.8$. And for atmospheric neutrino and K2K,
$|\Delta m^2_{atm}|\sim3\times10^{-3}~$eV$^2$ and
$\sin^22\theta_{atm}\sim1$.  Moreover LSND experiment~\cite{lsnd} may
have observed $\bar{\nu}_{\mu}\rightarrow\bar{\nu}_e$ transition which
can be also explained by neutrino oscillations with a large mass
scale, $|\Delta m^2_{LSND}|\sim(0.5-2.0)~$eV$^2$. Such results will
soon be checked by MiniBooNE~\cite{mini}.  These scales require four
neutrino oscillation framework (or three, if LSND results will not be
confirmed by MiniBooNE experiment) which imply, for UHE's of the order
1~EeV or higher, oscillation lengths much smaller than typical
distances from the sources of UHE neutrinos.  Consequently when
neutrino flavor oscillations are taken into consideration the flavor
proportion will be modified to
$\nu_e:\nu_{\mu}:\nu_{\tau}\sim1:1:1$. Therefore one expects a
considerable number of $\nu_{\tau}$ arriving at the Earth.

In this paper we investigate the possibility of detecting UHE cosmic
$\nu_{\tau}$ by means of a process in which a double Extensive Air
Shower (EAS) is identified, the so-called Double-Bang (DB) Phenomenon.
In that kind of event a $\nu_{\tau}$ interact with a quark via charged
current  creating one cascade of hadronic particles and a tau lepton
which subsequently decays producing a second cascade. DB Phenomenon
was first proposed for detectors where the neutrino energy should be
around $1~$PeV~\cite{learned}. It does not happen with neutrinos
different from $\nu_{\tau}$. The electron generated by an $\nu_e$
interacts immediately after being created and the muon generated by a
$\nu_{\mu}$, on the other hand, travel a much longer distance than the
size of the detector before interacting or even decaying. For energies
of the order of 1~EeV, where the radiative processes become more
important than ionization, the total energy loss in the atmosphere is
not important once that for those high energies, crossing
36000~g/cm$^2$ in iron, we estimate, by extrapolation, that the muon
will loose about 36\% of its initial energy~\cite{pdg}. So we may not
have DB events from them.

In order to identify a DB Phenomenon in the atmosphere, an optical
detector must be used to probe the longitudinal development of EAS's,
recording the fluorescence light emitted by the excited nitrogen
molecules of the Earth's atmosphere when the EAS passes through it.
One has to look for two EAS's coming from the same direction inside
the field of view (f.o.v.) of the detector, i.e., in the physical
space around the detector in which an event can be triggered. Based on
the phenomenology of the process we conclude that the features of
optical detectors like the Fluorescence Detectors (FD's) used by the
Pierre Auger Observatory~\cite{auger} favor the observation of DB
events with $\nu_{\tau}$ energies around 3~EeV. The estimated number
of DB events observed in these FD's varies from hundreds in a year to
few events in hundreds of years depending mainly on the primary
$\nu_{\tau}$ flux and cross section.

As we made the calculation based most on phenomenological aspects, the
conclusion of this work is not dependent on numerical simulations. To
have accurate conclusion based on simulations, first one needs to
improve the simulation programs including tau and $\nu_{\tau}$
interactions with particles in the atmosphere. There are some works
that study the differences between the longitudinal development of
EAS's generated by protons, heavier nuclei and different neutrino
flavors~\cite{jain,Ambrosio}. In reference~\cite{Ambrosio} the authors use
CORSIKA+Herwig Monte Carlo simulations to have $\nu_e$ and $\nu_{\mu}$
as primary particles, that will not produce DB's.

This paper is organized in the following way: Section~\ref{DBAuger}
has a brief introduction to the DB Phenomenon. Section~\ref{nbev}
describes how we calculate the DB event rate for a Pierre Auger-like
FD and for a configuration with 2 FD's and 90\% efficiency for
neutrino energy between 0.5 and 5~EeV.
Section~\ref{resdis} contains the number of events calculated for different
models and limits of UHE neutrino flux and discuss how could we take some
physical information from the two different efficiency approaches
described in Section~\ref{nbev}. Section~\ref{back} concerns possible
background events and the conclusions are in Section~\ref{conc}.

\section{\label{DBAuger}The Ultra-High Energy Double-Bang}
Studying the characteristics of FD's, such as its efficiency and
f.o.v. and the characteristics of the DB events generated by UHE
$\nu_{\tau}$, one can estimate the rate of this kind of event expected
in that kind of detector.

\begin{figure}
\resizebox{0.7\textwidth}{!}{
\includegraphics{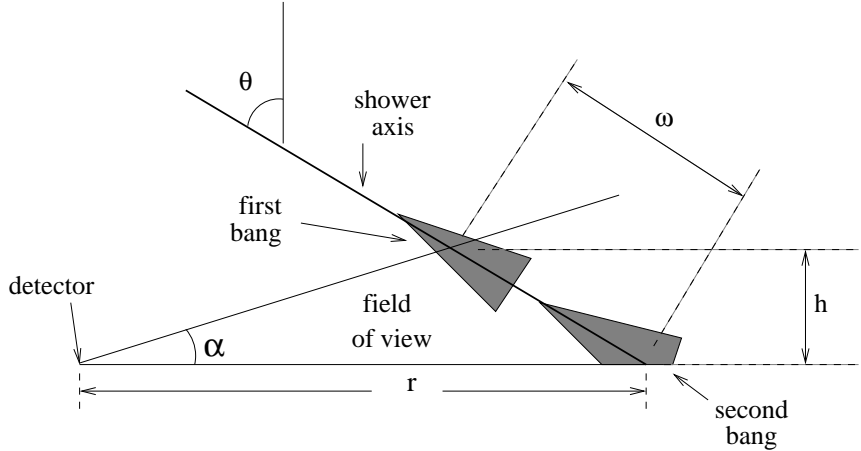} }
\caption{\label{esqdb}A schematic view of a Double-Bang and the f.o.v.
  of the Fluorescence Detector. See text in Section~\ref{DBAuger} for
  details.}
\end{figure}

Fig.~\ref{esqdb} shows a schematic view of an UHE DB with the detector
position and the time integrated development of the two EAS's, one
created by the UHE $\nu_{\tau}$ interacting with a nucleon in the
atmosphere and the other created by the decay of the tau generated in
the first interaction of the $\nu_{\tau}$. The f.o.v. of the Pierre
Auger Observatory FD, for example,  will be comprehended between
angles near the horizontal ($\alpha\sim2^o$) and $\alpha\sim30^o$, and
a maximum radius $r$ of approximately 30~km. The approximate maximal
height from where the DB can be triggered by the FD is h and $\omega$
is its projection in the DB propagation axis. The zenith angle is
represented by $\theta$.

The total amount of light emitted by the first EAS is related to
the energy transfered to the quark at the moment of the first
$\nu_{\tau}$  interaction, which we define as $E_1$. The neutrino
energy $E_{\nu}$ is the sum of the tau energy $E_{\tau}$ and $E_1$,
i. e., $E_{\nu}=E_1+E_{\tau}$. For charged current interactions above
0.1~EeV, approximately 20\% of the neutrino energy is transfered to
the quark~\cite{walker} and in our calculations we considered it
constant. The second EAS, resulting from the tau decay, carries an
energy $E_2$ of approximately $2/3~E_{\tau}$ and may be specially
visible when the tau decay is hadronic, which happens with a branching
ratio of around 63\%~\cite{pdg2}.

Therefore, very roughly, we have $\left<E_1\right>\sim1/5E_{\nu}$ and
$\left<E_2\right>\sim2/3\left<E_{\tau}\right>\approx8/15E_{\nu}$ and
the relation between $E_1$ and $E_2$ is given by:
$E_2/E_1\sim\frac{8}{15}E_{\nu}/\frac{1}{5}E_{\nu}\approx2.67$. The
distance traveled by the tau before decaying in laboratory frame is
$L={\gamma}ct_{\tau},$ where $\gamma=E_\tau/m_\tau$ and $t_{\tau}$ is
the tau mean life time, that has an error of approximately
0.4\%~\cite{pdg2}. When the tau energy is given in units of EeV,
$L\simeq  \frac{E_\tau}{\mbox{\scriptsize [EeV]}}\times 49~$km $\simeq
\frac{E_\nu}{\mbox{\scriptsize [EeV]}}\times 39.2~$km.

Now we compare the tau decay length with its attenuation length in
the Earth's atmosphere. The energy loss had been calculated
\cite{billoir} including bremsstrahlung, $e^+e^-$ par production and
deep inelastic scattering based on a model of the form
$-dE/dx=a+b(E)E$ where $a$ is the ionization energy loss and $b$ is
the sum of the other contributions due to radiative processes. The
second term, $b$, is dominant above a few 100~GeV. So we obtain an
attenuation length for the tau in the atmosphere $L_a=(\rho\sum
b)^{-1}\simeq33600~$km for $b=0.08\times10^{-7}$, $1.4\times10^{-7}$ and
$1.0\times10^{-7}$g$^{-1}$cm$^2$ from bremsstrahlung, par production
and deep inelastic scattering contributions respectively and
$\rho=1.2\times10^{-3}$g~cm$^{-3}$. The attenuation length ($L_a$) is
much longer than the decay length ($L$) and so we did not consider
energy loss for the tau propagation.

\section{\label{nbev}Event Rate}
To calculate the possible number of events in a FD, consider for
simplicity one Pierre Auger-like FD with a f.o.v. of 360$^o$. Then we
can write the equation for the DB event rate:
\begin{equation}\label{eq1}
\frac{dN_{events}}{dt}=\int_{E_{th}}^{\infty}
dE_{\nu}~\Phi_{\nu}(E_{\nu})~{\cal A}(E_{\nu},r,\theta)
\end{equation}
where, $E_{th}$ is the minimum detectable energy according to the
efficiency of the FD, $E_{\nu}$ is the $\nu_{\tau}$ energy,
$\Phi_{\nu}$ is the flux of UHE $\nu_{\tau}$ at the Earth depending on
the model of the extra galactic source of high-energy cosmic rays
considering maximal mixing and
\begin{equation}\label{eq:pdet}
{\cal A}(E_{\nu},r,\theta) = \int_{\Omega,A}d\Omega~dA
~P_{int}(E_{\nu},\theta)~F_{trig}(E_{\tau},r,\theta)~\Sigma(E_1,r)
\end{equation}
is the acceptance. $\Omega$ and $A$ are the solid angle covered by the
detector and the area under the f.o.v. of the detector respectively.
$P_{int}(E_{\nu},\theta)$ is the probability of the $\nu_{\tau}$ to
interact in a given point of the atmosphere,
$F_{trig}(E_{\tau},r,\theta)$ is a factor that indicates how probable
is to trigger a DB sign and $\Sigma(E_1,r)$ is the efficiency
of the FD. In the following subsections we explain each term accurately.
\subsection{Interaction Probability}
The interaction probability is given approximately by:
\begin{equation}\label{eq:pint}
P_{int}(E_{\nu},\theta) = \sigma_{CC}^{{\nu}N}(E_{\nu})~N_T(\chi)
\end{equation}
where $\sigma_{CC}^{{\nu}N}(E_{\nu})$ is the average charged current
cross section of the neutrino-nucleon interaction and $N_{T}(\chi)$,
the average total number of nucleons per squared centimeter at the
interaction point in the atmosphere. $N_{T}(\chi)=2N_A\chi(\theta)$,
where $N_A$ is the Avogadro's number and $\chi(\theta)$ is the
slant depth of the atmosphere within the points where the
neutrino must interact to generate a DB event inside the f.o.v.
of the FD.

\begin{figure}
\resizebox{0.7\textwidth}{!}{
\includegraphics{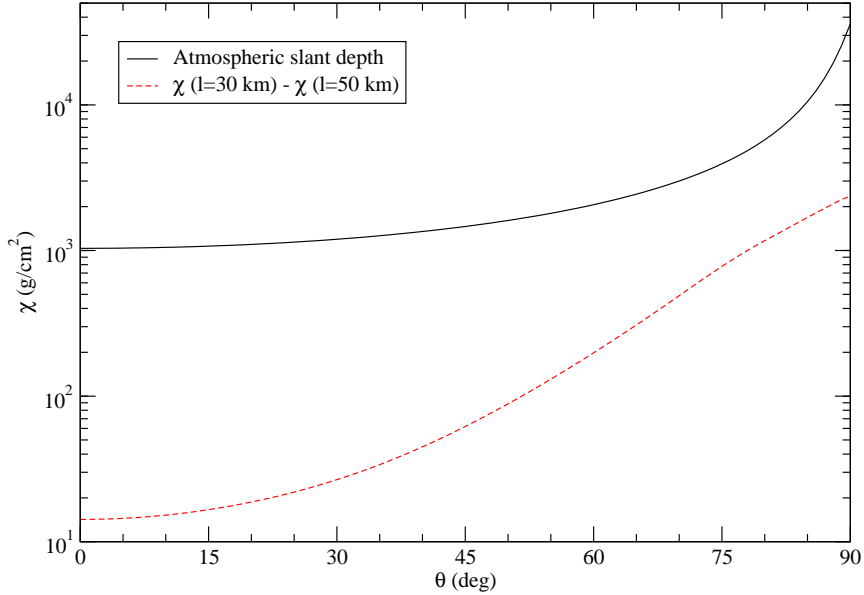}
}
\caption{\label{densxangul}Atmospheric slant
  depth and total depth within the region where UHE $\nu_{\tau}$'s have
  to interact to generate DB's, as a function of the zenith angle.}
\end{figure}

Considering the Earth's curvature, the atmospheric slant depth can be
approximately written as:
\begin{equation}\label{eq:prof}
\chi(\theta,l)=\int_\lambda\rho(H=l\cos\theta+\frac{(l\sin\theta)^2}{2R})
d\lambda
\end{equation}
where $\lambda$ is the path along the arrival direction from the source
until the interaction point in the atmosphere, $\rho$ is the
atmospheric density, $H$ the vertical height, $l$ is the distance
between the interaction point and the point toward the particle goes
through on Earth (the slant height), and $\theta$, the zenith
angle. The atmospheric depth as a function of the zenith angle is
shown in Fig.~\ref{densxangul} where we can see that the probability
for a neutrino to interact giving raise to a vertical EAS is very
low because the atmospheric depth is about 1000 g/cm$^2$,
approximately 36 times less than the atmospheric slant depth in the
horizontal case.


Taking the tau decay length $L(E_{\tau})=40~$km, for $E_{\nu}=1~$EeV, plus a
distance of 10~km for the second EAS to reach its maximum in the case
of a horizontal EAS~\cite{billoir}, we calculated the first
interaction occur approximately 50~km faraway from the detector. In
this case may be difficult to detect the maximum of the first EAS
because it will probably develop before reaching the f.o.v. of the
FD. We also estimate that if the first interaction happens about 30~km
faraway from the detector, the maximum of the first EAS can be seen,
but in the other hand may be difficult to detect the maximum of the
second EAS that probably will reach the ground before its maximum
development. To calculate the number of events presented in
Section~\ref{resdis} we considered the first interaction have to occur
in a point within 50~km and 30~km faraway from the detector.

The interaction probability is calculated with $\chi(\theta)$ given by:
\begin{equation}\label{eq:prodif}
\chi(\theta)=\int_{l_i}^{l_f}
\frac{\partial\chi(\theta,l)}{\partial l} dl =
\chi(\theta,l_f=30~ \mbox{km})-\chi(\theta,l_i=50~ \mbox{km})
\end{equation}

As the cross section of UHE neutrinos is unknown,
usually one adopts the extrapolation of parton distribution functions
and Standard Model (SM) parameters far beyond the reach of
experimental data. In this way, one can estimate a value for the cross
section of the neutrino-nucleon interaction of about
$10^{-32}$~cm$^2$, for energies around 1~EeV. Some authors say that
this extrapolation gives a neutrino-nucleon cross section that is too
high~\cite{weiler} but others use models that increase this same cross
section to typical hadronic cross section values~\cite{jain}. In this
work we use the following cross section parametrization:
\begin{equation}\label{eq:cs}
\sigma_{CC}^{\nu N}=(5.53+5.52)\times10^{-36}
\left(\frac{E_{\nu}}{\mbox{ [GeV]}}\right)^{0.363}\mbox{ cm}^2
\end{equation}
which is the SM extrapolation for the  neutrinos plus anti-neutrinos
and nuclei cross section in charged current interactions , which have
10\% accuracy within the energy range $10^{-2}<E($EeV$)<10^3$ when
compared with the results of the CTEQ4-DIS parton
distributions~\cite{gandhi}.
\subsection{Trigger Factor}
We define the trigger factor as given by:
\begin{equation}\label{eq:ptrig}
F_{trig}(E_{\tau},r,\theta)=P_{had}~P_L~\frac{\omega(r,\theta)}{L(E_{\tau})}
\end{equation}
where, $P_{had}$ is the hadronic branching ratio of tau decay
($P_{had}\simeq0.63$~\cite{pdg2}), $P_L$ is the mean percent amount of
taus that decay within the distance $L(E_{\tau})$ ($P_L\simeq0.63$),
$L(E_{\tau})$ is the distance traveled by the tau in laboratory frame
and $\omega(r,\theta)$, as can be seen in Fig.~\ref{esqdb}, is the
approximate size of the shower axis inside the f.o.v. of the detector
where the vertical plane containing the shower axis passes through the
center of the FD. In Eq.~\ref{eq:ptrig}, we are imposing
$\omega(r,\theta)/L(E_{\tau})=1$ if $\omega(r,\theta)>L(E_{\tau})$
so that we have a conservative estimation of the trigger factor. We
considered only showers moving away from the detector since, in the
opposite case, a large amount of \v{C}erenkov light arrives together
with the fluorescence light, spoiling a precise data
analysis~\cite{bellido}.
\subsection{Efficiency}
\begin{figure}
\resizebox{0.7\textwidth}{!}{
\includegraphics{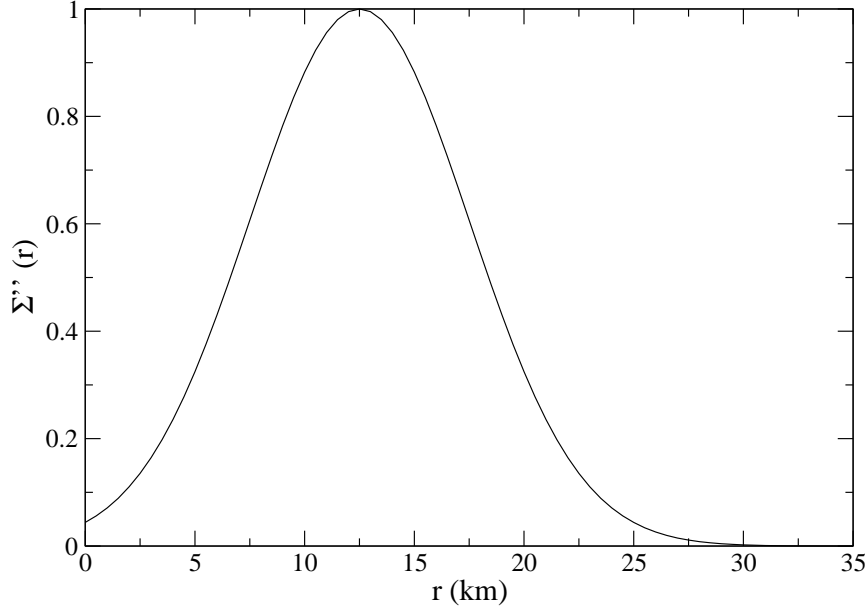}
}
\caption{\label{gauss}Efficiency as a function of the distance from
  the center of the detector.}
\end{figure}

\begin{figure}
\resizebox{0.7\textwidth}{!}{
\includegraphics{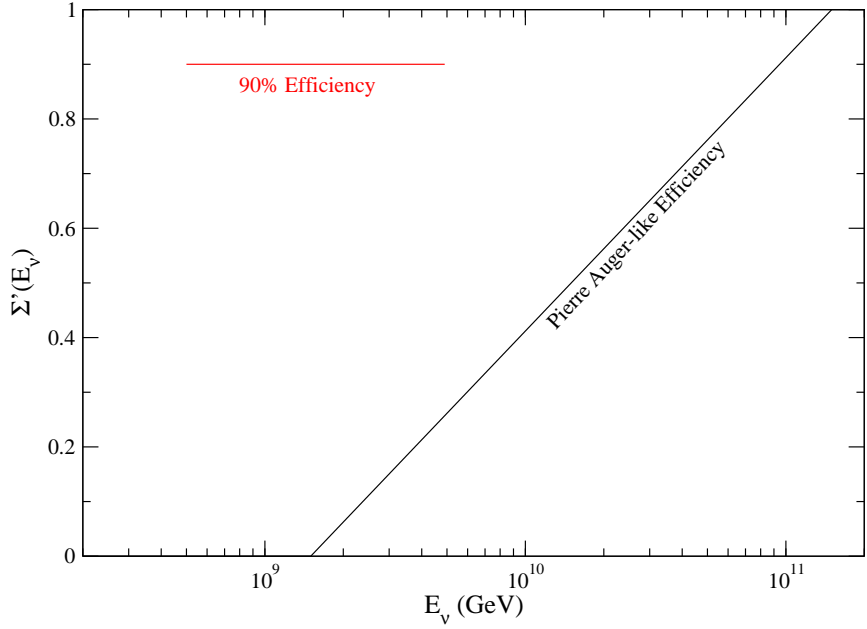}
}
\caption{\label{effic}Efficiency as a function of the neutrino
  energy. The Pierre Auger-like efficiency was used to calculate the
  event rate showed in columns $N_1$, $N_2$ and $N_3$ of Table~\ref{tab1},
  and the 90\% efficiency was used to calculate the number of events
  for two FD's with 60$^o$ f.o.v. showed in the last column of
  Table~\ref{tab1}.}
\end{figure}

The efficiency of the FD was estimated as:
\begin{equation}\label{eq:eff}
\Sigma(E_1,r)=\Upsilon~\Sigma'(E_1)~\Sigma''(r)
\end{equation}
where, $\Upsilon$ is the fraction of the time the fluorescence
detector will work ($\Upsilon\simeq0.1$ because the fluorescence
detector can only operate in clear moonless nights), $\Sigma'(E_1)$ is
the efficiency depending on the energy of the first EAS of the DB
Phenomenon that is less energetic than the second one and
$\Sigma''(r)$ is the efficiency depending on the distance from the
EAS core, where it reaches the ground, to the FD. $\Sigma'(E_1)$ and
$\Sigma''(r)$ depend on the characteristics of each detector. For
$\Sigma''(r)$ we used a Gaussian distribution centered at $r=12.5~$km faraway
from the detector and variance of 5.0~km. The behavior of
$\Sigma''(r)$ can be seen in Fig.~\ref{gauss}. We analyze two $\Sigma'(E_1)$
cases. For the first case we considered
$\Sigma'(E_1)$ rising logarithmically from 0 to 1 in the energy range
between approximately 0.3~EeV$~<E_1<~$30~EeV. This value may be coherent
with the characteristics of a Pierre Auger-like FD. In the second case we
considered 90\% efficiency ($\Sigma'(E_1)=0.9$) for neutrino energies between
0.5 and 5~EeV. The behavior of both $\Sigma'(E_1)$ can be seen in
Fig.~\ref{effic}, in terms of $E_{\nu}$.

\section{\label{resdis}Results}
Using Eq.~\ref{eq1} with all the phenomenological considerations given
above, we calculated the expected DB event rate which can be seen in
Table~\ref{tab1} for different models and limits of UHE cosmic ray
flux and in different energy intervals. The last column of
Table~\ref{tab1} shows the event rate in an hypothetical case with
90\% efficiency in the more relevant energy range for DB events (0.5
EeV $< E_{\nu} <$ 5 EeV), using 2 FD's with $\alpha=60^o$ (see
Fig.~\ref{esqdb}).
\begin{table}
\caption{\label{tab1}Number of events in the Pierre Auger-like FD
  during a period of one year, calculated in different regions of the
  energy spectrum and for different models and limits of cosmic ray
  flux. The last column was calculated with different considerations
  on the FD characteristics. See text for details. TD-92 stands for
  the model in reference~\cite{bhat}; TD-96 for the model in
  reference~\cite{sigl}; AGN-95J for the model in~\cite{karl}; WB
  for~\cite{wb} and MPR for~\cite{mpr}.}
\begin{ruledtabular}
\begin{tabular}{llll|l}
Models &
$N_1$\footnote{$E_{\nu}>1.0~$EeV} &
$N_2$\footnote{$E_{\nu}>10~$EeV} &
$N_3$\footnote{$E_{\nu}>100~$EeV} &
$N_4$\footnote{Two FD's with $\alpha=60^o$ and 90\% efficiency for
  energies 0.5 EeV $< E_{\nu} <$ 5 EeV.}\\
\hline
TD-92(0)    & 1.83                   & 0.46
                 & 0.03                   & 118              \\

TD-92(0.5)  & 0.03                   & 0.01
                 & 0.001                  & 1.46 \\

MPR         & 0.005                  & 9.0$\times10^{-4}$
                 & 3.7$\times10^{-5}$     & 0.48 \\

TD-92(1.0)  & 0.004                  & 0.002
                 & 3.2$\times10^{-4}$     & 0.093 \\

TD-92(1.5)  & 0.002                  & 7.1$\times10^{-4}$
                 & 1.3$\times10^{-4}$     & 0.037 \\

AGN-95J     & 6.1$\times10^{-4}$     & 1.1$\times10^{-4}$
                 & 4.7$\times10^{-6}$     & 0.060 \\

WB          & 1.1$\times10^{-4}$     & 2.0$\times10^{-5}$
                 & 8.4$\times10^{-7}$     & 0.011 \\

TD-96       & 4.7$\times10^{-8}$     & 4.8$\times10^{-9}$
                 & 8.3$\times10^{-11}$    & 9.0$\times10^{-6}$ \\
\end{tabular}
\end{ruledtabular}
\end{table}

\begin{figure}
\rotatebox{0}{ \resizebox{0.8\textwidth}{!}{
\includegraphics{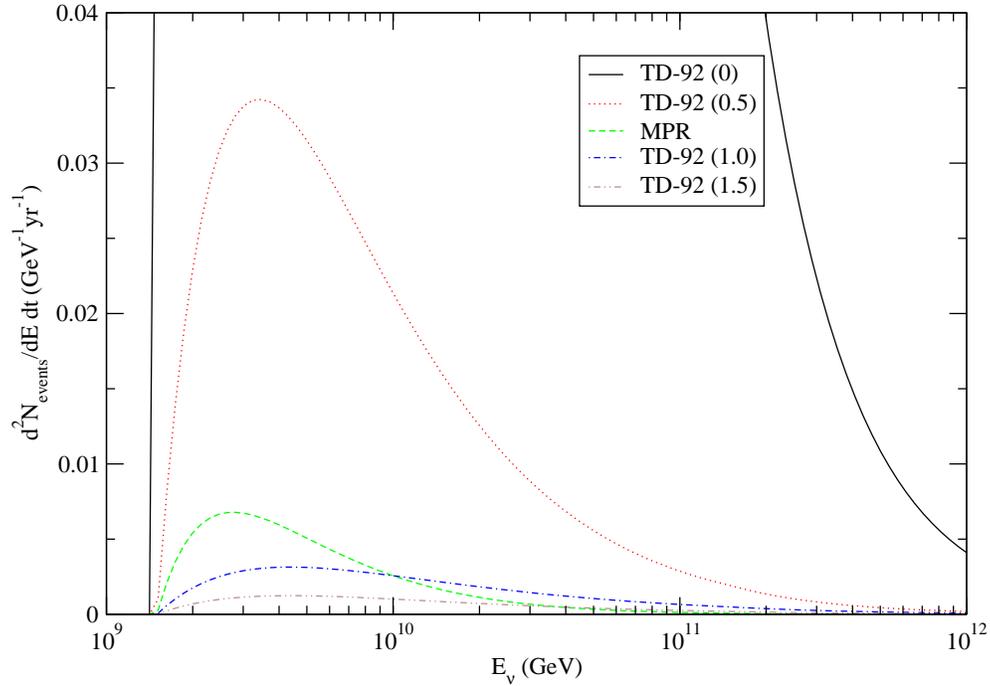} }
}
\caption{\label{nevents}Differential event rate for a Pierre Auger
  like FD for different models of cosmic ray flux.}
\end{figure}

From Table~\ref{tab1} and Fig.~\ref{nevents} one can learn which is
the energy interval which is  relevant to detect DB events with a
Pierre Auger-like FD. The
models WB~\cite{wb} and MPR~\cite{mpr} are limits for the UHE neutrino
flux based on cosmic ray observations. Both consider the neutrinos
coming from the interactions of protons and photons in the sources
generating pions that will decay into muons, electrons and
neutrinos. The basic difference is that the authors in WB state that
the sources are completely transparent to the protons and in the other
hand MPR say that the sources may have some opacity to the protons
that generate neutrinos in the interactions with the ambient light in
the source. So there could be some neutrino flux that arrive at the
earth but it might not be associated with the cosmic ray flux
observations. A reasonable flux model might predict an event rate
between these two limits.

\begin{figure}
\rotatebox{0}{ \resizebox{0.8\textwidth}{!}{
\includegraphics{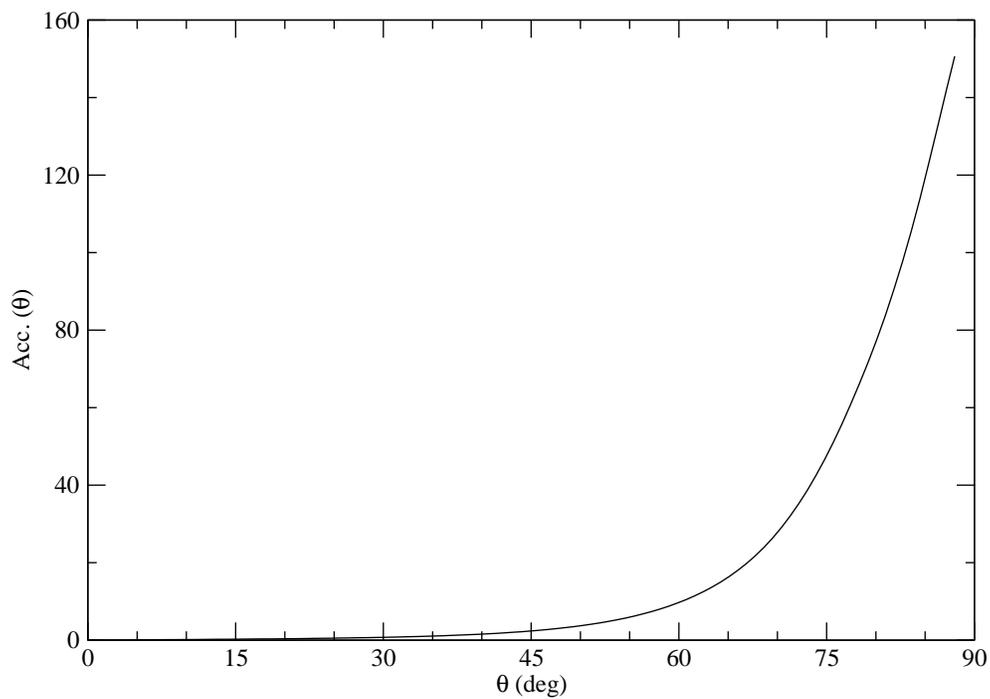} }
}
\caption{\label{accept}Dependence of the acceptance with the zenith angle.}
\end{figure}

Fig.~\ref{accept} shows the dependence of the acceptance with the
zenith angle. The acceptance is higher for events coming from almost
horizontal angles, but it is significant even for angles around 60 degrees.

Concerning the longitudinal development of the DB as a function of the
incident angle and energy of the primary neutrino, the convolution of
the terms $F_{trig}(E_{\tau},r,\theta)$ and $\Sigma(E_1,r)$
restricts the energies observed. For relatively low energies
($E_1<0.3~$EeV) the efficiency $\Sigma'(E_1)$ of the detector will be
low and for relatively high energies the factor
$F_{trig}(E_{\tau},r,\theta)$ may be too small because $L(E_{\tau})$
in Eq.~\ref{eq:ptrig} will be too large. In the hypothetical case of 90\%
efficiency for energies between 0.5 and 5~EeV and two FD's with
$\alpha=60^o$, a significant number of events can be measured.

\section{\label{back}Background events}
We consider here the possibility for a particle of the cosmic
radiation to masquerade a DB event depending on the accuracy of the
detector.  The probability for a proton, for example, to generate two
EAS's and masquerade the DB generated by a neutrino depends on two
possibilities: 1) that the primary proton interaction generates some
fragment that will give rise to a secondary shower deep in the
atmosphere with energy higher then the first. 2) that another shower
created by some independent particle interacts deep in the atmosphere
masquerading the second EAS of the DB.

In the possibility 1, the second EAS will be created by the decay or
interaction of the fragment deep in the atmosphere.  Usually the
primary proton generating an EAS looses roughly half of its energy to
the secondary particles that constitute the EAS and therefore it is
very hard that a possible second EAS has more energy than the first
one. There may be some cases where the proton looses only a few amount
of its initial energy and that a fragment of the EAS created by the
proton decays or interacts creating a second EAS with energy higher
than the first one. It could generate a background for DB events and
it has to be more carefully studied. Now, considering that for
energies of the order 1~EeV we have a cosmic ray flux of less than 1
particle per km$^2$ per year and that the only particles that could
probably interact deep in the atmosphere are neutrinos, generating the
second independent EAS near the detector, the chance that the primary
particle and this second independent neutrino come from the same solid
angle direction, in the same kilometer squared, interacting in a time
interval of the tau mean life time in the laboratory frame of $\gamma
t_{\tau}\approx 131\times \frac{E_\tau}{\mbox{\scriptsize
[EeV]}}~\mu$s is approximately 1 in 10$^7$, what exclude the
possibility 2. The direction of the two EAS's can be identified
specially if two FD's trigger the same DB event (with only one
detector, it must be difficult to know the direction of the EAS in the
plane that contains the EAS and the detector).

Based on this assumptions, $E_2/E_1$ may be a good parameter to
identify DB events if the measured energies from the two EAS's
are accurate enough. The error in the energy measured by
a FD depends mainly on the atmospheric conditions but hardly will
exceed 50\%. For a DB event the situation is optimistic because the
most important is the relation between the energies of the two EAS and
this error is smaller than the error of the absolute energy of an
ordinary EAS. We can make a conservative estimation of the error in
the average ratio $E_2/E_1$ considering the error in the absolute
energy of 50\%. This will give a relative error to the energy ratio of
70\%. So, since in average $E_2/E_1\approx2.67$ as deduced in
Section~\ref{DBAuger}, then considering such an error we find 95\% of
the events such that the energy ratio $E_2/E_1 > 1$. Then $E_2/E_1 >
1$ could be one minimal condition to identify DB events.

Because of the mean life time of the tau, we believe that it is
probable that relatively low energy DB will be superimposed looking
like an ordinary EAS, i.e., a single EAS generated by a proton. If one
detects an ordinary EAS profile of relatively high energy
($E_{EAS}>10$~EeV), that cannot be considered a DB event with ordinary
EAS profile because at such high energies the tau decay length would
separate the two EAS's of a DB Phenomenon. For relatively low energies
($E_{EAS}\sim0.1$~EeV) it must be considered the possibility of a DB
event with an ordinary EAS profile and it has to be more detailed
studied.

\section{\label{conc}Conclusion}

Taking into consideration neutrino oscillations, one expects that one
third of the high-energy neutrino flux arriving at the Earth should be
composed  of $\nu_{\tau}$. These neutrinos can interact in the Earth's
atmosphere generating a 
DB event. Many recent
works~\cite{CCPZ,Wilczy,Ave,billoir,aramo} have studied the potential
of the Pierre Auger Observatory to detect almost horizontal air
showers generated by 
UHE neutrinos. Here we
phenomenologically investigate the potential of a Pierre Auger-like
FD to observe DB events.

To calculate the DB event rate we considered different models and
limits of UHE neutrino flux in the case of full
mixing oscillations and the cross section for charged current
interactions only. We computed
the amount of matter the neutrino and the subsequent 
EAS's have to cross in the atmosphere which depends on
the incident angle of the neutrino and on tau properties as its decay
length when it carries 80\% of the incident neutrino energy and its hadronic
branching ratio. The 
neutrino interaction point depth was considered to be within
two extreme points where the DB Phenomenon could be detectable.
We also considered the detector geometry and trigger configuration
when we computed the field of view of the detector and
the efficiency depending on the energy and
distance. We have not considered the  energy dependence of the
distance where the efficiency is maximal.

DB events have very particular characteristics. Different from the
neutrino events in surface detectors, DB events do not need to come
from the very near-horizontal angles. Despite the low probability of
interacting at the top of the atmosphere, we can also have
$\nu_{\tau}$ creating DB events with incident angles 
of approximately 60$^o$ or larger. 
DB events also can have a lower primary neutrino energy, around
1~EeV, different from the energies around 50~EeV and beyond expected
for an ordinary EAS generated by the highest energy cosmic rays.

Some authors~\cite{wb,mpr} predict limits for the UHE neutrinos that give
very few DB events in a Pierre Auger-like FD. This is because the
energy range where the DB can be detected is very strict. For EAS
energies less then 0.3~EeV the efficiency of the FD detectors may be
too low and for $\nu_{\tau}$ energies greater then 20~EeV the two EAS's
are too separated. In the $\nu_{\tau}$ energy range between
approximately 2~EeV and 10~EeV a considerable part of the two EAS's
that characterize a DB may be detected by the FD's and then we could
have a DB trigger. Monte Carlo simulations with tau and $\nu_{\tau}$
must be made to study better some aspects as possible background
events and the expected features of this kind of phenomenon,
accounting, for example, for fluctuations in the EAS maximum that
would make the observation of DB events even more difficult.

Despite the fact the DB Phenomenon may be very rare, it is very
important to be prepared for its possible detection, specially in case
the Pierre Auger ground array detect near-horizontal air showers which
can indicate a sign for electron and/or muon neutrinos. Consequently
oscillations imply a considerable number of $\nu_{\tau}$ too. With
such a motivation, the Auger Observatory trigger could be calibrated
to be more sensitive for energies around 1~EeV.  With good efficiency
in this energy range, more detectors and more years collecting data we
could have more significant statistics. In Table~\ref{tab1}, we
presented numbers of DB event rate per year expected for a Pierre
Auger-like FD and also for an optimistic hypothetical case with 2
FD's, $\alpha=60^o$\footnote{The possibility of increasing the
overture of FD in Pierre Auger Observatory is presently under
consideration in the collaboration.} (see Fig.~\ref{esqdb}) and 90\%
efficiency for neutrino energies between 0.5 and 5~EeV. For the Pierre
Auger-like efficiency case, only the topological defect model TD-92(0)
predicts a significant number of DB events, of 1.83 per year for
neutrino energy bigger than 1~EeV.  On the other hand, assuming the
very feasible configuration with 2 FD's and 90\% efficiency for
neutrino energies between 0.5 and 5~EeV, models like TD-92(0),
TD-92(0.5) and also MPR limit can be tested predicting, respectively,
118, 1.46 and 0.48 events per year.

The potential of the DB Phenomenon to acquire valuable information
both in particle and astrophysics is irrefutable. For instance, the
cross section and flux of the ultra-high energy neutrinos are
speculative and can be investigated with DB events.

\begin{acknowledgments}
We thank Carlos Escobar, Vitor de Souza, Henrique Barbosa, Walter
Mello, Ricardo Sato, Gennaro Miele, Ofelia Pisanti, Dmitry Semikoz and
Guenter Sigl for valuable help and comments on the present work. This
research was partially supported by ``Conselho Nacional de
Desenvolvimento Cient\'{\i}fico e Tecnol\'ogico - CNPq'', the European
Union Programme of High Level Scholarships for Latin America -
Programme Al$\beta$an, scholarship no. E04D044701BR, the Spanish grant
BFM2002-00345 and ``Funda\c{c}\~ao de Amparo \`a Pesquisa do Estado de
S\~ao Paulo - FAPESP''
\end{acknowledgments}

\bibliography{astro_3}

\begin{thebibliography}{31}
\expandafter\ifx\csname natexlab\endcsname\relax\def\natexlab#1{#1}\fi
\expandafter\ifx\csname bibnamefont\endcsname\relax
  \def\bibnamefont#1{#1}\fi
\expandafter\ifx\csname bibfnamefont\endcsname\relax
  \def\bibfnamefont#1{#1}\fi
\expandafter\ifx\csname citenamefont\endcsname\relax
  \def\citenamefont#1{#1}\fi
\expandafter\ifx\csname url\endcsname\relax
  \def\url#1{\texttt{#1}}\fi
\expandafter\ifx\csname urlprefix\endcsname\relax\def\urlprefix{URL }\fi
\providecommand{\bibinfo}[2]{#2}
\providecommand{\eprint}[2][]{\url{#2}}

\bibitem[{\citenamefont{Greisen}(1966)}]{G}
\bibinfo{author}{\bibfnamefont{K.}~\bibnamefont{Greisen}},
  \bibinfo{journal}{Phys. Rev. Lett.} \textbf{\bibinfo{volume}{16}},
  \bibinfo{pages}{748} (\bibinfo{year}{1966}).

\bibitem[{\citenamefont{Zatsepin and Kuz'min}(1966)}]{ZK}
\bibinfo{author}{\bibfnamefont{G.~T.} \bibnamefont{Zatsepin}} \bibnamefont{and}
  \bibinfo{author}{\bibfnamefont{V.~A.} \bibnamefont{Kuz'min}},
  \bibinfo{journal}{JETP Letters} \textbf{\bibinfo{volume}{4}},
  \bibinfo{pages}{78} (\bibinfo{year}{1966}).

\bibitem[{\citenamefont{Halzen and Saltzberg}(1998)}]{HSaltzberg}
\bibinfo{author}{\bibfnamefont{F.}~\bibnamefont{Halzen}} \bibnamefont{and}
  \bibinfo{author}{\bibfnamefont{D.}~\bibnamefont{Saltzberg}},
  \bibinfo{journal}{Phys. Rev. Lett.} \textbf{\bibinfo{volume}{81}},
  \bibinfo{pages}{4305} (\bibinfo{year}{1998}).

\bibitem[{\citenamefont{Halzen}(1998)}]{Halzen}
\bibinfo{author}{\bibfnamefont{F.}~\bibnamefont{Halzen}},
  \emph{\bibinfo{title}{Lectures on neutrino astronomy: Theory and
  experiment}}, \bibinfo{howpublished}{Lectures presented at the TASI School}
  (\bibinfo{year}{1998}), \eprint{astro-ph/9810368}.

\bibitem[{\citenamefont{Learned and Pakvasa}(1995)}]{learned}
\bibinfo{author}{\bibfnamefont{J.~G.} \bibnamefont{Learned}} \bibnamefont{and}
  \bibinfo{author}{\bibfnamefont{S.}~\bibnamefont{Pakvasa}},
  \bibinfo{journal}{Astropart. Phys.} \textbf{\bibinfo{volume}{3}},
  \bibinfo{pages}{267} (\bibinfo{year}{1995}).

\bibitem[{\citenamefont{Holanda and Smirnov}(2002)}]{solar}
\bibinfo{author}{\bibfnamefont{P.~C.} \bibnamefont{Holanda}} \bibnamefont{and}
  \bibinfo{author}{\bibfnamefont{A.~Y.} \bibnamefont{Smirnov}},
  \bibinfo{journal}{Phys. Rev. D} \textbf{\bibinfo{volume}{66}},
  \bibinfo{pages}{113005} (\bibinfo{year}{2002}).

\bibitem[{\citenamefont{Fogli et~al.}(2001)\citenamefont{Fogli, Lisi, and
  Marrone}}]{atmospheric}
\bibinfo{author}{\bibfnamefont{G.~L.} \bibnamefont{Fogli}},
  \bibinfo{author}{\bibfnamefont{E.}~\bibnamefont{Lisi}}, \bibnamefont{and}
  \bibinfo{author}{\bibfnamefont{A.}~\bibnamefont{Marrone}},
  \bibinfo{journal}{Phys.Rev. D} \textbf{\bibinfo{volume}{64}},
  \bibinfo{pages}{093005} (\bibinfo{year}{2001}), \eprint{hep-ph/0105139}.

\bibitem[{\citenamefont{{Eguchi et al.}}(2003)}]{kamland}
\bibinfo{author}{\bibfnamefont{K.}~\bibnamefont{{Eguchi et al.}}},
  \bibinfo{journal}{Phys. Rev. Lett.} \textbf{\bibinfo{volume}{90}},
  \bibinfo{pages}{021802} (\bibinfo{year}{2003}).

\bibitem[{\citenamefont{{Ahn et al.}}(2001)}]{ktokplb}
\bibinfo{author}{\bibfnamefont{S.~H.} \bibnamefont{{Ahn et al.}}},
  \bibinfo{journal}{Phys. Lett. B} \textbf{\bibinfo{volume}{511}},
  \bibinfo{pages}{178} (\bibinfo{year}{2001}).

\bibitem[{\citenamefont{{Ahn et al.}}(2003)}]{ktokprl}
\bibinfo{author}{\bibfnamefont{S.~H.} \bibnamefont{{Ahn et al.}}},
  \bibinfo{journal}{Phys. Rev. Lett.} \textbf{\bibinfo{volume}{90}},
  \bibinfo{pages}{041801} (\bibinfo{year}{2003}).

\bibitem[{\citenamefont{{Athanassopoulos et al.}}(1995)}]{lsnd}
\bibinfo{author}{\bibfnamefont{C.}~\bibnamefont{{Athanassopoulos et al.}}},
  \bibinfo{journal}{Phys. Rev. Lett.} \textbf{\bibinfo{volume}{75}},
  \bibinfo{pages}{2650} (\bibinfo{year}{1995}).

\bibitem[{\citenamefont{Bazarko}(2002)}]{mini}
\bibinfo{author}{\bibfnamefont{A.}~\bibnamefont{Bazarko}}, in
  \emph{\bibinfo{booktitle}{Proceedings of the 31st International Conference on
  High Energy Physcis (ICHEP02)}} (\bibinfo{address}{Amsterdam},
  \bibinfo{year}{2002}), \eprint{hep-ex/0210020}.

\bibitem[{\citenamefont{{Eidelman et al.}}(2004{\natexlab{a}})}]{pdg}
\bibinfo{author}{\bibfnamefont{S.}~\bibnamefont{{Eidelman et al.}}},
  \bibinfo{journal}{Phys. Lett. B} \textbf{\bibinfo{volume}{592}},
  \bibinfo{pages}{251} (\bibinfo{year}{2004}{\natexlab{a}}),
  \bibinfo{note}{pdg}.

\bibitem[{aug(1997)}]{auger}
\emph{\bibinfo{title}{Pierre auger project design report}}
  (\bibinfo{year}{1997}),
  \bibinfo{note}{http://www.auger.org/admin/DesignReport/index.html}.

\bibitem[{\citenamefont{{Jain et al.}}(2002)}]{jain}
\bibinfo{author}{\bibfnamefont{A.}~\bibnamefont{{Jain et al.}}},
  \bibinfo{journal}{Int. J. Mod. Phys.} \textbf{\bibinfo{volume}{A17}},
  \bibinfo{pages}{533} (\bibinfo{year}{2002}), \eprint{hep-ph/0011310 v2}.

\bibitem[{\citenamefont{{Ambrosio et al.}}(2003)}]{Ambrosio}
\bibinfo{author}{\bibfnamefont{M.}~\bibnamefont{{Ambrosio et al.}}},
  \bibinfo{type}{Tech. Rep.} \bibinfo{number}{GAP-2003-013},
  \bibinfo{institution}{Auger Colaboration} (\bibinfo{year}{2003}),
  \bibinfo{note}{http://www.auger.org/admin-cgi-bin/woda/gap\_notes.pl/},
  \eprint{astro-ph/0302602}.

\bibitem[{\citenamefont{Quigg et~al.}(1986)\citenamefont{Quigg, Reno, and
  Walker}}]{walker}
\bibinfo{author}{\bibfnamefont{C.}~\bibnamefont{Quigg}},
  \bibinfo{author}{\bibfnamefont{M.~H.} \bibnamefont{Reno}}, \bibnamefont{and}
  \bibinfo{author}{\bibfnamefont{T.~P.} \bibnamefont{Walker}},
  \bibinfo{journal}{Phys. Rev. Lett.} \textbf{\bibinfo{volume}{57}},
  \bibinfo{pages}{774} (\bibinfo{year}{1986}).

\bibitem[{\citenamefont{{Eidelman et al.}}(2004{\natexlab{b}})}]{pdg2}
\bibinfo{author}{\bibfnamefont{S.}~\bibnamefont{{Eidelman et al.}}},
  \bibinfo{journal}{Phys. Lett. B} \textbf{\bibinfo{volume}{592}},
  \bibinfo{pages}{414} (\bibinfo{year}{2004}{\natexlab{b}}),
  \bibinfo{note}{pdg}.

\bibitem[{\citenamefont{{Bertou et al.}}(2002)}]{billoir}
\bibinfo{author}{\bibfnamefont{X.}~\bibnamefont{{Bertou et al.}}},
  \bibinfo{journal}{Astropart. Phys.} \textbf{\bibinfo{volume}{17}},
  \bibinfo{pages}{183} (\bibinfo{year}{2002}), \eprint{astro-ph/0104452}.

\bibitem[{\citenamefont{Kusenko and Weiler}(2002)}]{weiler}
\bibinfo{author}{\bibfnamefont{A.}~\bibnamefont{Kusenko}} \bibnamefont{and}
  \bibinfo{author}{\bibfnamefont{T.}~\bibnamefont{Weiler}},
  \bibinfo{journal}{Phys.Rev.Lett.} \textbf{\bibinfo{volume}{88}},
  \bibinfo{pages}{161101} (\bibinfo{year}{2002}), \eprint{hep-ph/0106071 v2}.

\bibitem[{\citenamefont{Gandhi et~al.}(1998)\citenamefont{Gandhi, Quigg, Reno,
  and Sarcevic}}]{gandhi}
\bibinfo{author}{\bibfnamefont{R.}~\bibnamefont{Gandhi}},
  \bibinfo{author}{\bibfnamefont{C.}~\bibnamefont{Quigg}},
  \bibinfo{author}{\bibfnamefont{M.~H.} \bibnamefont{Reno}}, \bibnamefont{and}
  \bibinfo{author}{\bibfnamefont{I.}~\bibnamefont{Sarcevic}},
  \bibinfo{journal}{Phys. Rev. D} \textbf{\bibinfo{volume}{58}},
  \bibinfo{pages}{093009} (\bibinfo{year}{1998}).

\bibitem[{\citenamefont{Bellido}(1998)}]{bellido}
\bibinfo{author}{\bibfnamefont{J.~A.} \bibnamefont{Bellido}},
  \bibinfo{type}{Tech. Rep.} \bibinfo{number}{GAP-98-027},
  \bibinfo{institution}{Auger Colaboration} (\bibinfo{year}{1998}),
  \bibinfo{note}{http://www.auger.org/admin-cgi-bin/woda/gap\_notes.pl/}.

\bibitem[{\citenamefont{Bhattacharjee et~al.}(1992)\citenamefont{Bhattacharjee,
  Hill, and Schramm}}]{bhat}
\bibinfo{author}{\bibfnamefont{P.}~\bibnamefont{Bhattacharjee}},
  \bibinfo{author}{\bibfnamefont{C.~T.} \bibnamefont{Hill}}, \bibnamefont{and}
  \bibinfo{author}{\bibfnamefont{D.~N.} \bibnamefont{Schramm}},
  \bibinfo{journal}{Phys. Rev. Lett.} \textbf{\bibinfo{volume}{69}},
  \bibinfo{pages}{567} (\bibinfo{year}{1992}).

\bibitem[{\citenamefont{Sigl}(1996)}]{sigl}
\bibinfo{author}{\bibfnamefont{G.}~\bibnamefont{Sigl}}, \bibinfo{journal}{Space
  Science Reviews.} \textbf{\bibinfo{volume}{75}}, \bibinfo{pages}{375}
  (\bibinfo{year}{1996}).

\bibitem[{\citenamefont{Mannheim}(1995)}]{karl}
\bibinfo{author}{\bibfnamefont{K.}~\bibnamefont{Mannheim}},
  \bibinfo{journal}{Astropart. Phys.} \textbf{\bibinfo{volume}{3}},
  \bibinfo{pages}{295} (\bibinfo{year}{1995}).

\bibitem[{\citenamefont{Bahcall and Waxman}(2001)}]{wb}
\bibinfo{author}{\bibfnamefont{J.}~\bibnamefont{Bahcall}} \bibnamefont{and}
  \bibinfo{author}{\bibfnamefont{E.}~\bibnamefont{Waxman}},
  \bibinfo{journal}{Phys. Rev. D} \textbf{\bibinfo{volume}{64}},
  \bibinfo{pages}{023002} (\bibinfo{year}{2001}), \eprint{hep-ph/9902383 v2}.

\bibitem[{\citenamefont{Rachen et~al.}(2001)\citenamefont{Rachen, Protheroe,
  and Mannheim}}]{mpr}
\bibinfo{author}{\bibfnamefont{J.~P.} \bibnamefont{Rachen}},
  \bibinfo{author}{\bibfnamefont{R.~J.} \bibnamefont{Protheroe}},
  \bibnamefont{and} \bibinfo{author}{\bibfnamefont{K.}~\bibnamefont{Mannheim}},
  \bibinfo{journal}{Phys. Rev. D} \textbf{\bibinfo{volume}{63}},
  \bibinfo{pages}{023003} (\bibinfo{year}{2001}), \eprint{astro-ph/9812398 v3}.

\bibitem[{\citenamefont{{Capelle et al.}}(1998)}]{CCPZ}
\bibinfo{author}{\bibfnamefont{K.~S.} \bibnamefont{{Capelle et al.}}},
  \bibinfo{journal}{Astropart. Phys.} \textbf{\bibinfo{volume}{3}},
  \bibinfo{pages}{321} (\bibinfo{year}{1998}), \eprint{astro-ph/9801313}.

\bibitem[{\citenamefont{Wilczy\'{n}ski}(2000)}]{Wilczy}
\bibinfo{author}{\bibfnamefont{H.}~\bibnamefont{Wilczy\'{n}ski}},
  \bibinfo{type}{Tech. Rep.} \bibinfo{number}{GAP-00-020},
  \bibinfo{institution}{Auger Colaboration} (\bibinfo{year}{2000}),
  \bibinfo{note}{http://www.auger.org/admin-cgi-bin/woda/gap\_notes.pl/}.

\bibitem[{\citenamefont{{Ave et al.}}(2000)}]{Ave}
\bibinfo{author}{\bibfnamefont{M.}~\bibnamefont{{Ave et al.}}},
  \bibinfo{journal}{Astropart. Phys.} \textbf{\bibinfo{volume}{14}},
  \bibinfo{pages}{109} (\bibinfo{year}{2000}), \eprint{astro-ph/0003011}.

\bibitem[{\citenamefont{{Aramo et al.}}(2005)}]{aramo}
\bibinfo{author}{\bibfnamefont{C.}~\bibnamefont{{Aramo et al.}}},
  \bibinfo{journal}{Astropart. Phys.} \textbf{\bibinfo{volume}{23}},
  \bibinfo{pages}{65} (\bibinfo{year}{2005}), \eprint{astro-ph/0407638}.

\end{thebibliography}

\end{document}